\documentclass[aps,pre,reprint,showpacs,superscriptaddress]{revtex4-1}
\usepackage{graphicx,color,epsfig}
\usepackage{amssymb,amsmath}

\begin{document}
\title{Global interactions, information flow, and chaos synchronization}
\author{G. Paredes}
\affiliation{LFAC, Universidad Nacional Experimental del T\'achira, San Crist\'obal, Venezuela.}
\author{O. Alvarez-Llamoza}
\affiliation{Departamento de F\'isica, FACYT, Universidad de Carabobo, Valencia, Venezuela.}
\author{M. G. Cosenza} 
\affiliation{Centro de F\'isica Fundamental, Universidad de los
Andes, M\'erida, Venezuela.}
\date{Phys. Rev. E \textbf{88}, 042920 (2013)}

\begin{abstract}
We investigate the relationship between 
the emergence of chaos synchronization and the information flow 
in dynamical systems possessing homogeneous or heterogeneous global interactions whose origin can be external (driven systems) or internal (autonomous systems). 
By employing general models of coupled chaotic maps for such systems, 
we show that the presence of a homogeneous global field, either external or internal, for all times is not indispensable for achieving  complete or generalized synchronization in a system of chaotic elements. Complete synchronization can also appear with heterogeneous global fields; it does not requires the simultaneous sharing of the field by all the elements in a system.
We use the normalized mutual information and the information transfer between global and local variables to characterize complete and generalized synchronization.
We show that these information measures can characterize both types of synchronized states and also allow to discern the origin of a global interaction field. A synchronization state emerges when a sufficient amount of 
information provided by a field is shared by all the elements in the system, on the average over long times. 
Thus, the maximum value of 
the top-down information transfer
can be used as a predictor of synchronization 
in a system, as a parameter is varied.
 
\end{abstract}
\maketitle

\section{Introduction}
Global interactions in a system occur when all its elements are subject to a common influence or field. 
Global interactions appear naturally
in the description of many physical, biological and social systems, such as coupled oscillators \cite{Kuramoto,Nakagawa}, 
Josephson junction arrays \cite{Hadley}, charge density waves \cite{Gruner}, multimode lasers \cite{Wie}, parallel electric circuits, neural dynamics, ecological systems, evolution models \cite{Kaneko1}, economic exchange \cite{Yakovenko}, social networks \cite{Newman}, 
mass media models \cite{Media}, cross-cultural interactions \cite{Plos}, etc. 
A global interaction field may consist of an external environment acting 
on the elements, as in a driven dynamical system; or it may originate from the interactions between the elements, 
in which case, we talk of autonomous dynamical systems. In many cases, global interaction fields coexist with local
or short-range interactions.

Although systems with global interactions possess a simple topological connectivity structure 
--a fully connected network--,
they can exhibit a variety of collective behaviors, such as chaos synchronization,
dynamical clusters, nontrivial collective behavior, chaotic itineracy \cite{Kaneko1,Manrubia}, quorum sensing \cite{Ojalvo}, etc. 
These behaviors have been studied in models of globally coupled maps \cite{Kaneko2} and 
have been experimentally investigated in globally coupled oscillators in chemical, physical and biological systems \cite{Wang,DeMonte,Taylor,Roy}. 

In particular, chaos synchronization is a fundamental phenomenon in dynamical systems \cite{Pecora,Pikovsky}. Its investigation has provided insights into many natural processes and motivation for practical applications such as secure communications and control of nonlinear systems \cite{Bocaletti,Argyris,Uchida}. Complete 
synchronization in a system of dynamical elements subject to a global interaction field, either
external or autonomous, occurs when the state variables of all the elements and the global field converge to a single orbit in phase space. 
Generalized chaos synchronization, originally discovered in driven chaotic systems, arises when
all the state variables of the elements in the system get synchronized into an
orbit that is different from that of the drive \cite{Rulkov,Abarbanel}. The
concept of generalized synchronization of chaos has also been extended to the context of autonomous systems \cite{We}. This means that the chaotic state variables 
in a dynamical system can synchronize to each other but not to a 
coupling function containing information from those variables.

The occurrence of both forms of chaos synchronization in driven and in autonomous systems with global interactions
suggests that the nature, either
external or endogenous, of the global field acting on the elements in a system is irrelevant.  
At the local level, each element in the system is subject to a field that eventually induces some form of synchronization between that field and the element. 
In general, the local dynamics in systems with global interactions can be seen as a single drive-response system \cite{Parra,Manrubia}.  
In particular, if the time evolution of an external global field acting on a system is identical to that of an autonomous global field acting on a replica system, the corresponding local drive-response dynamics in both systems should be indistinguishable, and therefore the corresponding synchronized states are equivalent; i. e., they occur for the same parameter values in both systems \cite{JPCS}.

In many systems it is important not only to detect synchronized or other collective states, but also 
to understand the 
relationships between global and local scales that lead to such behaviors.  
For example, it has recently been argued that top-down causation
--where information flows from higher levels to lower levels in complex systems-- may be
a major contributor to evolutionary transitions and to the emergence of behaviors in living systems \cite{Davies},
and synchronization in neural systems has been described as a top-down information processing driven by a stimulus \cite{Falta}.

The above results suggest that the emergence of collective behaviors, 
such as a synchronized state, in a 
system is associated with the reception by its elements of some amount of information provided by a source, either external or endogenous to the system.
In this article we investigate 
the relationship between information flow between the global and local variables, and the emergence of
complete and generalized synchronization of chaos 
in dynamical networks with global interactions of different types. 
We employ information measures \cite{Shanon,Schreiber} that have been widely applied to quantify 
drive-response causal relationships between subsystems and interdependences between data sets
in many fields of science, including
linguistics \cite{Linguis}, electroencephalographic signals \cite{Palus}, neuroscience \cite{Neuro},
communication systems \cite{Comm}, dynamical systems \cite{Grebogi}, and climate networks \cite{Kurths}.
We show that these information measures can characterize complete and generalized synchronized states and also allow us 
to discern the origin, either external or endogenous, of a global interaction field. A given synchronization state emerges when a sufficient amount of the information transmitted by a field is shared by all the elements in the system, on the average over long times. Thus, the maximum value of the top-down
information transfer can be employed as a predictor of synchronization as a parameter in the system,
such as the coupling strength to the field, is varied. 

In Sec.~II we present a general coupled map model for systems with external or endogenous global interactions. 
and define the quantities to characterize synchronized states and information flow in such systems. Homogeneous global interaction fields, 
which may act intermittently, are considered in Sec.~III.  We extend the concept of a global field to 
include heterogeneous global interactions in Sec.~IV. Section~V contains the conclusions of this work. 

\section{Global interaction fields}
We describe a global interaction in a system as a field that can influence all the elements in the system. 
As a simple model of a dynamical system subject to a global interaction, we consider a system of $N$ coupled maps of the form
\begin{equation}
\label{sys}
\begin{array}{ll}
x^i_{t+1} = & w(x^i_t, y_t)  \\
y_{t+1}=& \phi(y_t,x_t^j),
\end{array}
\end{equation}
where $x^i_t$ ($i=1,2,\ldots,N$) represents the state variable of the $i$th map in the system at discrete time $t$, 
$y_t$ is a global interaction field that can affect each map at time $t$, and
$j \in Q$ where $Q$ is a subset of elements in the system.
Equation~(\ref{sys}) describes a system of elements interacting with a common dynamical environment that can receive feedback from the system. 
For simplicity, we shall focus on the presence of global interactions and will not include local interactions.

An external global field $y_t$ possesses its own dynamics, 
independent from the dynamics of the elements, given by
\begin{equation}
\label{Efield}
\phi(y_t,x_t^j)=g(y_t).
\end{equation}
On the other hand, an internal global field $y_t$ can be represented by 
\begin{equation}
\label{Gfield}
\phi(y_t,x_t^j)=h(x_t^j \, |\, j \in Q_t),
\end{equation}
where $h$ is a function of the states of a given subset $Q_t$ of elements in the system at time $t$. The coupling function $h$ may represent a constraint or a conservation law on the system.

\begin{figure}[h]
\begin{center}
\includegraphics[scale=0.32,angle=0]{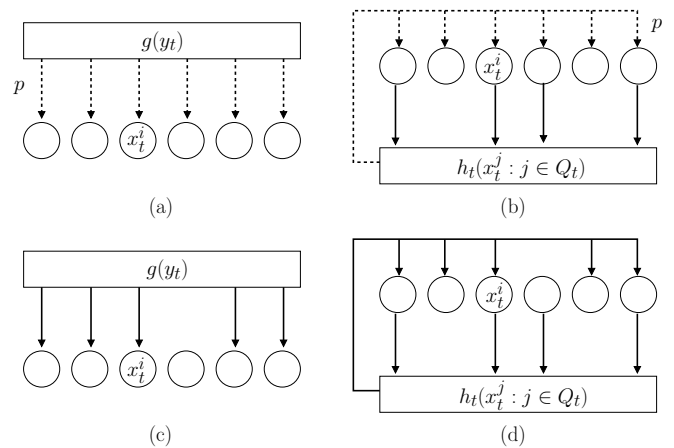}
\end{center}
\caption{Top panels: homogeneous global interactions. (a) External field $g(y_t)$ acting 
with probability $p$ on all elements. 
(b) Internal field $h( x^j_t \, |\, j \in Q_t)$ acting with probability $p$ on all elements. 
Bottom panels: heterogeneous global interactions. 
(c) External field $g(y_t)$ acting on a fraction $p$ of elements chosen at random
at every time.
(d) Internal field $h( x^j_t \, |\, j \in Q_t)$ 
acting on a fraction $p$ of elements chosen at random at every time.}
\label{f1}
\end{figure}

We shall consider the coupling of the maps to the global interaction field in the diffusive form
\begin{equation}
\label{diff}
w(x^i_t, y_t)=(1-\epsilon)f(x^i_t)+ \epsilon \phi(y_t,x_t^j) ,
\end{equation}
where $f$ describes the local dynamics of the maps, and the parameter $\epsilon$ is the strength of the coupling to the global
field. 
Since we are particularly interested in chaos synchronization, we choose for the local dynamics the logistic map 
$f(x_t^i)=4x_t^i(1-x_t^i)$, 
so that $f(x_t^i)$ is fully chaotic for $x_t^i \in [0,1]$.
In this paper, we consider both, driven and autonomous systems, subject to global interactions, whose schemes are 
illustrated in Fig.~\ref{f1}. 

\subsection{Synchronization states}

Synchronization in the system Eq.~(\ref{sys}) at a time $t$ corresponds to a state $x_t^i=x_t^j$, $\forall \, i,j$. Thus, 
a synchronized state can be described by the condition
$x_t^i=\bar x_t$, $\forall \, i$,
where  $\bar x_t$ is the instantaneous mean field of the system,
\begin{equation}
\label{mean}
\bar x_t=\frac{1}{N} \sum^N_{i=1} x^i_t \, .
\end{equation}

To characterize the occurrence of synchronization, we shall consider the asymptotic time-average
$\langle\sigma\rangle$ (after discarding a number of transients) of the instantaneous standard deviations
$\sigma_t$ of the distribution of state variables $x^i_t$, defined as
\begin{equation}
\label{sigma}
\sigma_t=\left[ \frac{1}{N} \sum_{i=1}^N \left( x^i_t - \bar x_t \right)^2 \right]^{1/2}.
\end{equation}
A synchronization state corresponds to $\langle \sigma \rangle=0$. In addition, we define the
asymptotic time-average
$\langle \delta \rangle$ (after discarding a number of transients) of the quantity
\begin{equation}
\label{delta}
\delta_t = | \bar x_t - y_t |.
\end{equation}

Two forms of synchronization can take place in the system  Eq.~(\ref{sys}) in relation to the global field $y_t$:
(i) \textit{complete synchronization}, given by the condition $x_t^i=\bar x_t=y_t$, i.e., all elements are 
synchronized to the field, and characterized by $\langle\sigma \rangle=0$ and $\langle \delta \rangle=0$; 
and (ii)
\textit{generalized synchronization}, corresponding to the condition $x_t^i=\bar x_t \neq y_t$, i.e., all elements 
are synchronized to each other but not the field, and described by $\langle\sigma \rangle=0$ and 
$\langle \delta \rangle \neq 0$. 
It has been shown that both types of synchronization can occur in systems with global interactions, 
for either autonomous or driven systems \cite{We}.  In this paper we shall use the numerical criteria $\langle \sigma \rangle < 10^{-7}$ and $\langle \delta \rangle < 10^{-7}$ for characterizing the zero values of these quantities.

In order to characterize the information exchange between the global field and the local dynamics in the system, 
we consider the following quantities: 
\renewcommand{\labelenumi}{(\arabic{enumi})}
\begin{enumerate}
 \item the normalized mutual information between two variables $y_{t}$ and $x_t$, based on Shanon's 
 mutual information \cite{Shanon},
\begin{equation}
\label{mutual}
M_{y,x}=-\frac{\sum_{x_t,y_t} P \left(x_{t},y_{t} \right) 
\log \left (\dfrac{ P \left(x_{t},y_{t} \right )}{P\left ( x_{t} \right) P\left ( y_{t} \right )} \right)}
{\sum_{x_t}P(x_t) \log P(x_t)} ;
\end{equation}
\item  the information transfer from a variable $y_{t}$ to a variable $x_t$, defined as \cite{Schreiber}
\begin{equation}
\label{trasn1}
T_{y,x}=\sum_{x_{t+1},x_{t},y_{t}} P \left(x_{t+1},x_{t},y_{t} \right)
\log\left(\frac{P\left(x_{t+1},x_{t},y_{t}\right) P\left( x_{t} \right) }{P\left( x_{t},y_{t} 
\right)P\left(x_{t+1},x_{t} \right)} \right),
\end{equation}
\end{enumerate}
where $P(x_{t})$ means the probability distribution of the time series of the variable $x_t$, $P(x_t,y_t)$ is the
joint probability distribution of $x_t$ and $y_t$, and so on. 
The quantity $M_{y,x}$ measures the overlap of the information content of the variables $y_t$ and
$x_t$; it represents how much the uncertainty about $x_t$ decreases if $y_t$ is known. 
The quantity $T_{y,x}$
measures the degree of dependence of $x_t$ on the variable $y_t$; i.e., the information required to represent
the value $x_{t+1}$ from the knowledge of $y_t$. Note that the information transfer is nonsymmetrical, i.e., 
$T_{y, x}\neq T_{x, y}$.
The normalized mutual information $M_{y,x}$ is symmetrical, i.e. $M_{y,x}=M_{x,y}$, and does not indicate the direction of the flow of information
between two interacting dynamical variables, as $T_{y, x}$ does.
When the two variables are synchronized, $x_t=y_t$. Then we obtain $M_{y,x}=1$ and  $T_{y, x}=0$.

\section{Homogeneous global interactions}
We describe a  homogeneous global interaction as a field shared simultaneously by all the elements in a system. Since, 
in general, the interaction with the field may not occur for all times, we consider a coupled map system subject to 
a homogeneous, intermittent, global interaction of the form
\begin{equation}
\label{int}
\forall i, \;  x^i_{t+1} = \left\lbrace 
\begin{array}{ll}
w(x^i_t, y_t), &    \text{with probability $p$},\\
f(x_t^i),  &   \text{with probability $(1-p)$} .
\end{array}
\right. 
\end{equation}

Each map in the system  
Eq.~(\ref{int}) is subject to the presence (or absence) of the same influence at any time. 
Then, the occurrence of complete or generalized 
synchronization between a local map and the global field  $y_t$ implies 
the same form of synchronization between the mean field of the system $\bar x_t$ and  $y_t$,
regardless of the nature, either external or endogenous, of the global field   $y_t$.

A system subject to a homogenous external field [Fig.~\ref{f1}(a)], corresponds to
\begin{equation}
\label{int_drive}
\begin{array}{ll}
\forall i, & x_{t+1}^i = \begin{cases}
(1-\epsilon)f(x^i_t)+ \epsilon g(y_t), \;  \text{with probability $p$}\\
f(x_t^i),   \quad \text{with probability $(1-p)$},
          \end{cases} \\
& y_{t+1} =  g(y_t).
\end{array}
\end{equation}
The auxiliary system approach introduced in Ref.~\cite{Abarbanel} implies that a driven map can
synchronize on identical orbits with another, identically driven map. 
The system Eq.~(\ref{int_drive}) can be regarded as one of multiple realizations for different initial conditions of a single, 
intermittently driven map. Thus, by extension, the elements in this system should synchronize 
with the external field in the same form as a single local map driven by that field does. 

A complete synchronized state in the system Eq.~(\ref{int_drive}) is given by $x_t^i=\bar x_t=y_t$, $\forall \. i$, 
and it can occur when the external field is equal to the local dynamics, $g=f$.  If $g \neq f$, generalized
synchronization, characterized by the condition $x_t^i = \bar x_t \neq y_t$, $\forall \, i$, may also arise in this system.

On the other hand, a system subject to an autonomous homogeneous global field [Fig.~\ref{f1}(b)] can be described as
\begin{equation}
\label{auto_syst}
\begin{array}{ll}
\forall i, &   x^i_{t+1} =  \begin{cases} 
(1-\epsilon)f(x^i_t)+ \epsilon  h(x_t^j\, |\, j \in Q_t), \text{with probability $p$},\\
f(x_t^i),  \quad   \text{with probability $(1-p)$} , 
         \end{cases} 
\end{array}
\end{equation}
where $Q_t$ is a subset consisting of $q \leq N$ elements of the system that may be chosen at random at each time $t$. 
Each map receives the same input from the endogenous global field $y_t=h$ at any $t$ with probability $p$.
Complete synchronization in the system Eq.~(\ref{auto_syst}) occurs when $f(x_t^i)=f(\bar x_t)=h$; 
while generalized synchronization appears if $f(x_t^i)=f(\bar x_t) \neq h$, $\forall \,  i$.

\subsection{Complete synchronization}
As examples of complete chaotic synchronization in systems having 
homogeneous global interactions, we consider the
driven system Eq.~(\ref{int_drive}) with $g=f$, and the autonomous system Eq.~(\ref{auto_syst}) subject to
a partial mean field coupling function defined as
\begin{equation}
\label{partial}
h( x^j_t\, |\,j \in Q_t)=\frac{1}{q} \sum^q_{j=1} f(x^j_t),
\end{equation}
where $q \leq N$ maps are randomly chosen at each time $t$. 
For these systems, the condition $\langle \delta \rangle=0$ implies $\langle \sigma \rangle =0$ and, therefore, complete synchronization.

\begin{figure}[h]
\begin{center}
\includegraphics[scale=0.75,angle=0]{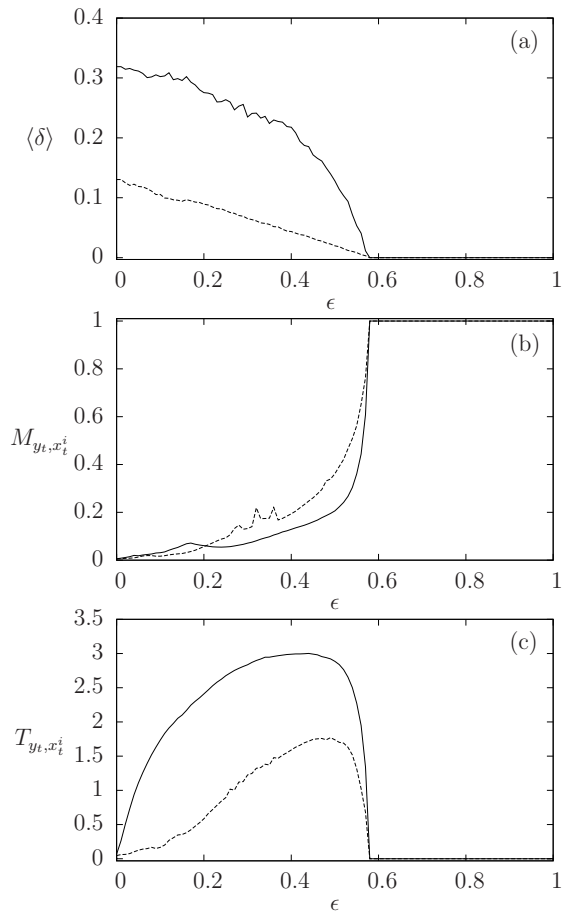} 
\end{center}
\caption{Complete chaos synchronization in systems with homogeneous global fields. 
(a) $\langle\delta \rangle$ vs. $\epsilon$, 
(b) mutual information $M_{y_t,x_t^i}$  vs. $\epsilon$, and 
(c) information transfer $T_{y_t,x_t^i}$ vs. $\epsilon$.
On each panel, the continuous line corresponds to the homogeneous driven system, Eq.~(\ref{int_drive}) with $g=f$
and the dashed line corresponds to the  homogeneous autonomous system,  Eqs.~(\ref{auto_syst}) and (\ref{partial}).
Both information measures are calculated with $2 \times 10^5$ points in the time series, after discarding transients, and
averaged over 50 randomly chosen maps. The number of states used
to calculate the corresponding probability distributions is $100$. 
The same conditions are used in Figs.~\ref{f3} and \ref{f5}.
Fixed parameters: $p=0.8$, $N=10^{4}$, $q/N=0.4$.}
\label{f2}
\end{figure}

Figure~\ref{f2}(a) shows the quantity $\langle \delta \rangle$ as a function of the coupling parameter $\epsilon$, for both the homogeneous driven system and the homogeneous autonomous system,  with fixed values of $p$ and $q/N$.  
Complete synchronization for both systems takes place at a critical value $\epsilon_c=0.579$, for which  
$\langle \delta \rangle <10^{-7}$. 

Figures~\ref{f2}(b) and \ref{f2}(c) show, respectively, the normalized mutual information $M_{y_t,x_t^i}$ and the information transfer $T_{y_t,x_t^i}$ between the homogeneous global field and one map, averaged over 50 randomly chosen maps, for both systems as functions of $\epsilon$. These averaged quantities give practically the same result
as for just one randomly chosen map.
The results shown are also independent of $q$ for large enough system system size $N$. 
We observe that, as the coupling strength $\epsilon$ increases, the global field and the local variables become more correlated, and the normalized mutual information for both systems increases until $M_{y_t,x_t^i}=1$ at the value $\epsilon_c$. In the complete synchronization region,  $\epsilon\geq \epsilon_c$, we find the constant values $M_{y_t,x_t^i}=1$ and  $T_{y_t,x_t^i}=0$ for both systems, signaling complete synchronization in each case. Once complete chaos synchronization is established, the evolution of the global field, regardless of its source, is identical to that of the maps. Thus, the mutual or the transfer information cannot distinguish between the driven and the autonomous systems in a regime of complete synchronization. 

On the other hand, just before vanishing at the critical value $\epsilon_c$, the information transfer for both systems becomes maximum. 
This indicates that, as the critical values of the parameters for the onset of complete chaos synchronization are approached, the flow of information from the global field to the local maps must be large. 
Figure~\ref{f2}(c) shows that the maximum value of the information transfer for the driven system is greater than the corresponding maximum value for the autonomous system. Thus, in the vicinity of parameter values for the emergence of complete synchronization, an autonomous global field needs to convey less information to the local maps than an external driving field. This suggests that the  information transfer $T_{y_t,x_t^i}$ can serve as a predictor of a state of complete synchronization in the parameter space of driven and autonomous systems with homogenous global interactions. Moreover, this quantity can distinguish between these two types of systems near the onset of complete synchronization.

\subsection{Generalized synchronization}
If the functional form of the global field is different from that of the local dynamics, generalized synchronization
may occur in a system subject to a homogeneous global interaction. For example, 
consider an external field in a driven system Eq.~(\ref{int_drive}) such as
\begin{equation} 
\label{GSdrive}
g(y_t)= \frac{\mu}{2}\left( 1-\left|2 y_t-1 \right|\right) ,
\end{equation}
with $\mu=1.98$ and $y_t \in [0,1]$. Then at synchronization in the driven system Eqs.~(\ref{int_drive})-(\ref{GSdrive}), 
we have $x_t^i=\bar x_t \neq y_t$. 
Similarly, in an autonomous system, Eqs.~(\ref{auto_syst}), consider a homogeneous global interaction different from a mean field, such as the coupling function
\begin{equation}
\label{shift}
 h(x^j_t\, |\,j \in Q_t)= \frac{\mu}{2}\left[ 1-\left| 2 \, \left( \frac{1}{q} \sum_{j=1}^q x^j_t\right) 
-1 \right|\right] ,  
\end{equation}
with $\mu=1.98$, where $q \leq N$ elements are chosen at random at each time $t$. Then, a synchronized state in 
the autonomous system Eqs.~(\ref{auto_syst})-(\ref{shift}) corresponds to
$f(x_t^i)=f(\bar x_t) \neq h$. 
For the fields chosen above, the functional form of the autonomous field in a synchronized state 
is  similar to that of the drive, $h=g(\bar x_t)$. However, the time evolution of $h$ at synchronization is not
necessarily identical to that of $g(y_t)$. 

\begin{figure}[h]
\begin{center}
\includegraphics[scale=0.75,angle=0]{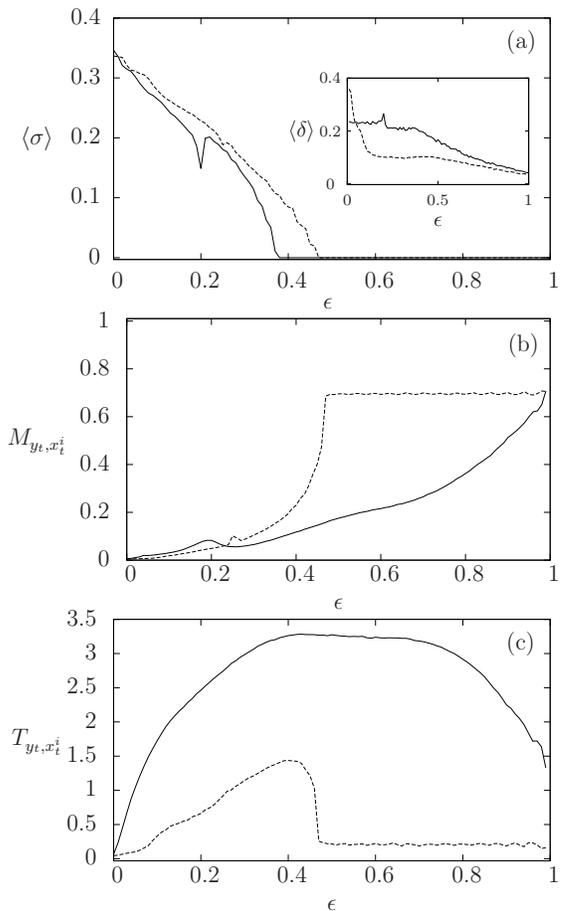}
\end{center}
\caption{Generalized chaos synchronization in systems with homogeneous global fields. 
(a) $\langle\sigma \rangle$ vs. $\epsilon$ (inset: $\langle\delta \rangle$ vs. $\epsilon$);
(b) $M_{y_t,x_t^i}$  vs. $\epsilon$; and 
(c) $T_{y_t,x_t^i}$ vs. $\epsilon$.
On each panel,  the continuous line corresponds to a homogeneously driven system Eq.~(\ref{int_drive}) with $g$ given in Eq.~(\ref{GSdrive}) and the dashed line corresponds to the
homogeneous autonomous system Eqs.~(\ref{auto_syst})-(\ref{shift}). 
Fixed parameters are $p=0.8$, $N=10^{4}$, $q/N=0.4$.}
\label{f3}
\end{figure}

Figure \ref{f3}(a) shows the quantity  $\langle \sigma \rangle$ as a function 
of the coupling parameter $\epsilon$ for both systems with  homogeneous global interactions, the driven system with $g$ given by Eq.~(\ref{GSdrive})
and the autonomous system with $h$ given by Eq.~(\ref{shift}).   
These systems get synchronized at different values of $\epsilon$ for which $\langle \sigma \rangle<10^{-7}$.
The inset in Fig.~\ref{f3}(a) shows that the quantity $\langle \delta \rangle$ for both systems does not vanish when $\epsilon$ is varied, indicating that the synchronized state in both cases corresponds to generalized synchronization.

Figure~\ref{f3}(b) shows $M_{y_t,x_t^i}$ for both systems, 
as a function of $\epsilon$. 
In contrast to the constant value  $M_{y_t,x_t^i}=1$ exhibited by the normalized mutual information
for both systems in a state of complete synchronization [Fig.~\ref{f2}(b)], the behavior of $M_{y_t,x_t^i}$ 
in the regime of generalized 
synchronization is different for each system. The normalized mutual information for the autonomous system in a generalized synchronized state reaches an almost constant value, $M_{y_t,x_t^i}=0.695 <1$, since the time series of the local maps and the coupling function $h$ are not identical. For the driven system, $M_{y_t,x_t^i}$ increases monotonically with increasing $\epsilon$, but the values of $M_{y_t,x_t^i}$ are below the value of this quantity for the autonomous system in the region of generalized synchronization.
Therefore, in a generalized synchronization state, the amount of information shared between the field $h$ and the local maps in the autonomous system is greater than that between the external field $g$ and the maps in the driven system. 

Figure~\ref{f3}(c) shows the
information transfer $T_{y_t,x_t^i}$ versus $\epsilon$ for both systems. 
Similarly to the behavior observed for complete synchronization, as the coupling strength approaches the critical 
value  $\epsilon_c$ for the emergence of generalized synchronization, the information transfer in 
the autonomous system becomes maximum. Also, the values of $T_{y_t,x_t^i}$ for the driven system are greater than the  values of this quantity for the autonomous system. 
However, in the generalized chaos synchronization regime, for $\epsilon > \epsilon_c$,
the information transfer in both systems does not vanish; and
the values of  $T_{y_t,x_t^i}$ for the driven system are greater than the values of this quantity for the autonomous system. This means that the autonomous field must provide less information to the local maps than an external drive for sustaining generalized synchronization. This behavior should be expected since the autonomous field $h$
already contains information about the dynamics of the elements in the system.
At the onset of generalized synchronization, both $T_{y_t,x_t^i}$ and $M_{y_t,x_t^i}$ for the driven system are continuous while they are discontinuous for the autonomous system.
Thus, 
the quantities $M_{y_t,x_t^i}$ and $T_{y_t,x_t^i}$ can distinguish between the driven and the autonomous systems in a state of generalized synchronization, in contrast to the case of complete synchronization.

\subsection{Dynamics at the local level}
At the local level in a system with a homogeneous global field, each element is subject to a field that eventually induces some form of synchronization between that element and the field, similarly to a single master-slave system. 
Thus, the local dynamics can be seen as a single drive-response map system where a drive $g$ acts with probability $p$ on a map $f$.
In particular, the linear stability analysis of the complete synchronized state for 
the single driven-map 
yields the condition \cite{We}
\begin{equation}
p \ln \vert 1-\epsilon \vert +\lambda_f <0,
\end{equation}
where $\lambda_f$ is the Lyapunov exponent of the map $f$. A stable completely synchronized state 
occurs when this condition is fulfilled. On the other hand, a stable generalized synchronized state in both kinds of homogeneous system 
can be numerically determined with the criterion $\langle \sigma \rangle < 10<{-7}$
on the space of parameters $(p,\epsilon)$.

\begin{figure}[h]
\begin{center}
\includegraphics[scale=0.55,angle=0]{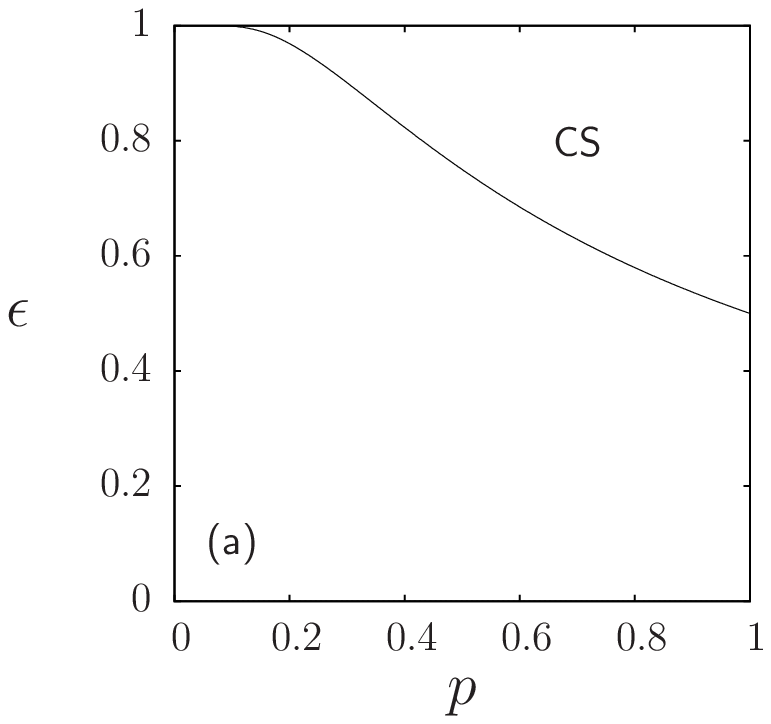} 
\includegraphics[scale=0.55,angle=0]{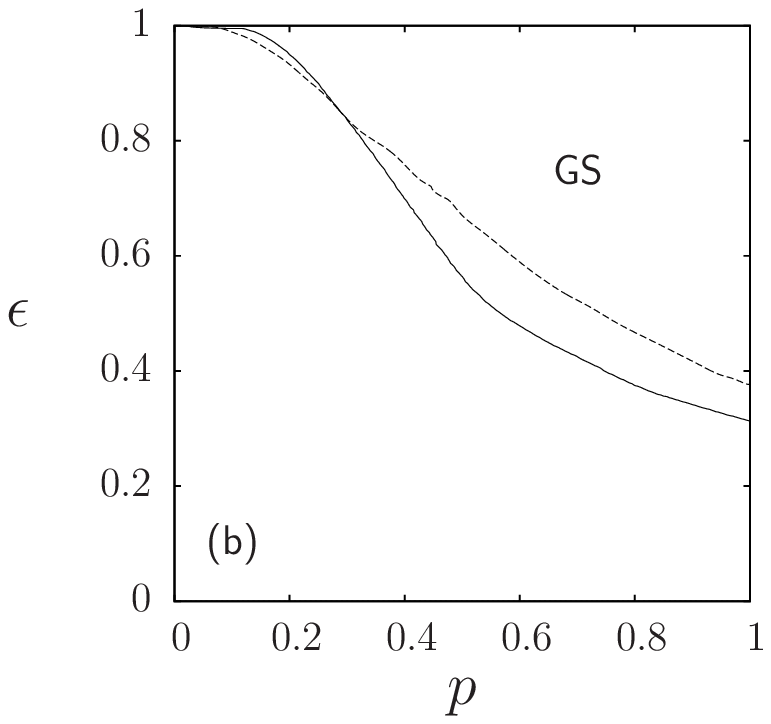} 
\end{center}
\caption{Regions for chaos synchronization on the plane $(p,\epsilon)$ for systems with homogeneous global interactions. 
(a) Complete synchronization (CS) for both the 
homogeneous driven system, Eq.~(\ref{int_drive}) with $g=f=4x(1-x)$, and the homogeneous autonomous system,  Eqs.~(\ref{auto_syst}) and (\ref{partial}). The boundary of the region where complete synchronization takes place
is given by $\epsilon = 1-e^{-\lambda_f/p}$, with $\lambda_f=\ln2$ for the map $f$.  
(b) Generalized synchronization (GS) for both the homogeneously driven system Eqs.~(\ref{int_drive}) and (\ref{GSdrive}) (continuous line), and for the homogeneous autonomous system Eqs.~(\ref{auto_syst}) and (\ref{shift}) (dashed line),
with $N=10^{4}$, $q/N=0.4$.}
\label{f4}
\end{figure}

Figure~\ref{f4} shows the regions where complete and generalized synchronization can be found on the plane $(p,\epsilon)$ for the systems with homogeneous global interactions considered here. 
The region of parameters for complete  synchronization is the same for both the autonomous and the driven systems.
The regions corresponding to generalized synchronization are not identical for these systems with the chosen functional forms of their global fields.

\section{Heterogeneous global interactions} 
The concept of a global field can be extended beyond the concept of spatial homogeneity. 
In this respect, we consider a system with heterogeneous global 
interactions, as follows
\begin{equation}
\label{hetero}
 x^i_{t+1} = \left\lbrace 
\begin{array}{ll}
w(x^i_t, y_t), &    \text{if $i \in R_t$},\\
f(x_t^i),  &   \text{if $i \notin R_t$} .
\end{array}
\right. 
\end{equation}
where $R_t$ is a subset containing $pN$ elements of the system, with $p \leq 1$, which may be chosen at random at each time $t$.
 Thus, the average fraction of elements  coupled to the field
in 
Eq.~(\ref{hetero}) at any given time is $p$, so that not all the maps in the system
receive the same influence at all times.   
In comparison, the coupling of the elements to the field in 
systems with homogeneous global interactions, Eq.~(\ref{int}), is simultaneous and uniform; each map receives the same
influence from the field $y_t$ at any $t$ with probability $p$. At the local level, each map in the system 
with heterogeneous global interactions Eq.~(\ref{hetero})
is subject, on the average, to the 
 global field $y_t$ with probability $p$ over long times.
For $p=1$, the homogeneous system Eq.~(\ref{int}) 
and the heterogeneous system Eq.~(\ref{hetero}) are identical. 

In the case of 
an external field (Fig.~\ref{f1}(c)), Eq.~(\ref{hetero}) takes the form
\begin{equation}
\label{part_drive}
\begin{array}{l}
x_{t+1}^i = \begin{cases}
(1-\epsilon)f(x^i_t)+ \epsilon g(y_t), \;  \text{if $i \in R_t$}\\
f(x_t^i),  \quad  \text{if $i \notin R_t$},
          \end{cases} \\
y_{t+1} =  g(y_t).
\end{array}
\end{equation}

For 
an autonomous field [Fig.~\ref{f1}(d)], the coupled map system Eq.~(\ref{hetero}) becomes
\begin{equation}
x^i_{t+1} = \left\lbrace 
\begin{array}{ll}
(1-\epsilon)f(x^i_t)+ \epsilon h(x^j_t:j \in Q_t), \;  \text{if $i \in R_t$},\\
f(x_t^i), \quad  \text{if $i \notin R_t$},
\end{array}
\right. 
\label{pauto_syst}
\end{equation}
where, again, $Q_t$ is a subset consisting of $q \leq N$ elements of the system that 
may be chosen at random at each time $t$. Each map in Eq.~(\ref{pauto_syst}) is subject, on the average, 
to the same coupling function $h$ with probability $p$ over long times. The same  condition holds for each
map with respect to the drive $g$ in the heterogeneously driven system Eq.~(\ref{part_drive}). Then, if $g$ exhibits the same temporal evolution as $h$, the synchronization behavior of the autonomous system Eq.~(\ref{pauto_syst}) should 
be similar to the behavior of the driven system Eq.~(\ref{part_drive}) over long times. 

\subsection{Complete synchronization}

When the heterogeneously driven system Eq.~(\ref{part_drive}) gets synchronized, we have $x_t^i= \bar x_t$. However, the  
synchronized solution exists only if $g=f$. Therefore, only complete synchronization $x_t^i= \bar x_t=y_t$ can take place
in this system. On the other hand, a synchronized state in the heterogeneous autonomous system Eq.~(\ref{pauto_syst}) occurs when $f(x_t^i)=f(\bar x_t)$. However, this synchronized solution exists only if $h=f(\bar x_t)$. Therefore, as in the case of the heterogeneous driven system, only complete synchronization, where $f(x_t^i)=f(\bar x_t)=h$, can emerge in the heterogeneous autonomous system Eq.~(\ref{pauto_syst}). As an example of a coupling function $h( x^j_t:j \in Q_t)$ leading to complete synchronization in the heterogeneous autonomous system Eq.~(\ref{pauto_syst}), we choose the partial mean field Eq.~(\ref{partial}).  

\begin{figure}[h]
\begin{center}
\includegraphics[scale=0.75,angle=0]{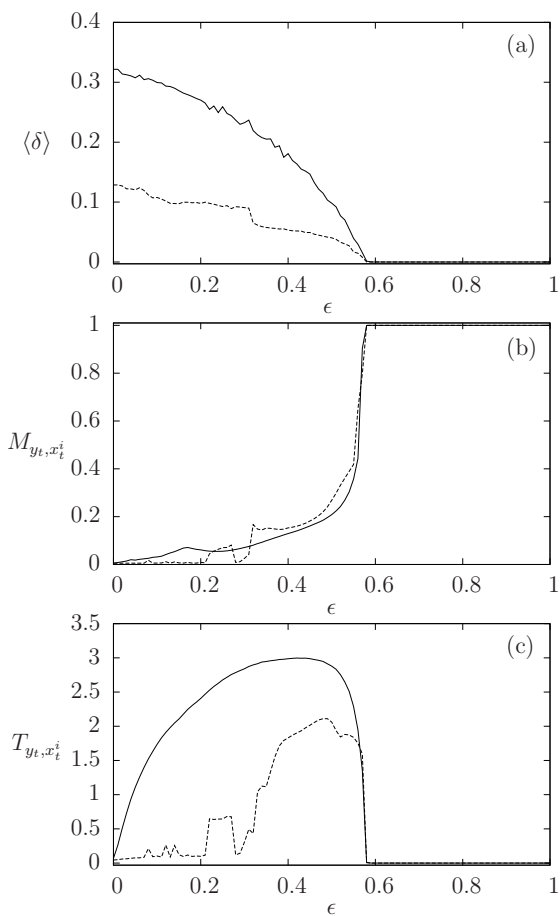}
\end{center}
\caption{Complete chaos synchronization in systems with heterogeneous global fields. 
(a) $\langle\delta \rangle$ vs. $\epsilon$,   
(b) $M_{y_t,x_t^i}$  vs. $\epsilon$; and 
(c) $T_{y_t,x_t^i}$ vs. $\epsilon$.
On each panel, the continuous line corresponds to the heterogeneously driven system Eq.~(\ref{part_drive})
with $g=f$, and the dashed line corresponds to the heterogeneous autonomous system Eqs.~(\ref{pauto_syst}) and (\ref{partial}).
Fixed parameters: $p=0.8$, $N=10^{4}$, $q/N=0.4$.}
\label{f5}
\end{figure}

Figures~\ref{f5}(a)-\ref{f5}(c) show the quantities $\langle\delta \rangle$, $M_{y_t,x_t^i}$ and $T_{y_t,x_t^i}$, respectively, as functions of  $\epsilon$ for both heterogeneous systems, driven and autonomous, with global interactions. Both systems reach complete  chaos synchronization at the critical value  $\epsilon_c=0.579$. 

The information transfer in Fig.~\ref{f5}(c) becomes maximal 
previous to the synchronization threshold, similarly to the behavior observed in homogeneous systems. 
Thus, a maximum in the information transfer $T_{y_t,x_t^i}$ in the space of parameters can be regarded
as a precursor to a state of synchronization, either complete or generalized. Figures~\ref{f2}(c), \ref{f3}(c), and \ref{f5}(c) reveal that a lesser amount of information flow from the global field to the local maps is necessary for the emergence of synchronization in autonomous systems, in comparison to that required for synchronization in driven systems 
possessing similar functional forms of their global fields and identical parameter values. 

In either homogeneous or heterogeneous autonomous systems, complete synchronization occurs 
independently of the number $q$ of elements randomly chosen in the function 
$h$, or if the $q$ chosen elements are always the same. Thus, the reinjection of an autonomous coupling function $h$, although containing partial information about the system,  to a fraction of randomly selected elements suffices to achieve complete synchronization. 
If the elements in subset $R_t$ receiving the coupling function $h$ or the drive $g$ are always the same, then only elements in this subset reach complete synchronization, since only those elements share the same information, on the average. 

\section{Conclusions}
We have investigated  the relationship between the emergence of synchronization and the information flow 
in dynamical systems possessing global interactions. We have used the normalized mutual information $M_{y_t,x_t^i}$ and the information transfer $T_{y_t,x_t^i}$ between global and local variables to characterize complete and generalized synchronization in models of coupled chaotic maps for such systems.

We have found that the presence of a homogeneous global field, either external or internal, for all times is not indispensable for achieving  complete or generalized synchronization in a system of chaotic elements. Complete synchronization can also appear with heterogeneous global fields; it does not requires the simultaneous sharing of a global field by all the elements in the system. Furthermore, the global coupling function in autonomous systems does not need to depend on all the internal variables for reaching synchronization and, in particular, its functional form is not determinantal for generalized synchronization. 

In both systems with homogeneous or heterogeneous 
global fields, at the local level each element 
is subject, on the average, to a field that eventually induces some form of synchronization between that element
and the field, similarly to a single drive-response system.
Then, a set of elements identical to the response and subject to a global field 
that behaves as the drive also synchronizes in a similar manner.

What becomes essential for the emergence of a given synchronization state is that all the elements in the system share 
a sufficient amount of information provided by a field, on the average, over time. This amount is characterized by the maximum value of the information transfer $T_{y_t,x_t^i}$ 
previous to the critical values of parameters for either complete or generalized synchronization.
Therefore, the quantity $T_{y_t,x_t^i}$ could be employed to anticipate the occurrence of a  state of synchronization in the space of parameters of a system possessing a global interaction field. 
Furthermore, the form in which information flows from macroscopic to microscopic scales for the emergence of synchronization, as measured by the quantities $M_{y_t,x_t^i}$ and  $T_{y_t,x_t^i}$,
differs between a driven and an autonomous system with global interactions, even if they have similar functional forms for their local dynamics or for their global fields.  In summary, we have found that (i) near the onset of complete synchronization when a parameter is varied, the maximum of the information transfer $T_{y_t,x_t^i}$ for a driven system is greater than that for an autonomous system; 
(ii) near the onset of generalized synchronization,  
the normalized mutual information $M_{y_t,x_t^i}$ and $T_{y_t,x_t^i}$
exhibit sharp changes for an autonomous system, while these quantities exhibit a smooth behavior for a driven system; and 
(iii)  in a state of generalized synchronization, 
$T_{y_t,x_t^i}$ is greater for a driven system than for an autonomous system and
$M_{y_t,x_t^i}$ is smaller for a driven system
than for an autonomous system. 

Our results suggest that these information measures could be used to characterize, and possibly also to predict, other forms of collective behaviors observed in dynamical systems having global interactions. 
Further extensions of this work include the investigation of the relationship between top-down information flow between global and local scales, and the emergence of collective behaviors
and structures in more complex dynamical networks.  

\section*{Acknowledgments}
This work was supported by Project No. C-1827-13-05-B from CDCHTA, Universidad de Los Andes, Venezuela. 
M. G. C. is grateful to the Senior Associates Program of 
the Abdus Salam International Centre for Theoretical Physics, Trieste, Italy.

\end{document}